# Analysis of Logistic Map for Pseudorandom Number Generation in Game Development

Chenxiao Zhou*

*Li Po Chun United World College of Hong Kong, Shatin, New Territories, Hong Kong

Email: chen25@lpcuwc.edu.hk

## ABSTRACT

Many popular video games use pseudorandom number generators to create randomly distributed locations for game objects as highly unpredictable as possible. Some scenarios like game competition also need "reproducible randomness", namely the random results can be reproducible if given the same seed input. Existing random generation methods have limited choices for seed input. To address this limitation, this study analyzes a chaotic map called the Logistic Map for game development. After analyzing the properties of this chaotic map, I developed a pseudorandom sequence generation algorithm and a generation algorithm of random locations of game objects. Experiments on the game of Snake demonstrate that the Logistic Map is viable for game development. The reproducible randomness is also realized with the proposed algorithm.

## INTRODUCTION

Randomness exists everywhere in nature. However, true randomness is surprisingly challenging to simulate on a computer. Over the years, programmers created something called the pseudorandom number generator (PRNG) [4] by feeding a constantly changing seed to a function that outputs what appears to be statistically random results. Many popular video games use PRNGs to create randomly distributed game objects. Examples include the pieces in Tetris, bombs in Minesweeper, and food in Snake [7] - just to name a few. PRNGs intend to create random locations for the game objects that are **highly unpredictable** as possible. The unpredictability will bring challenges to game players and thus attract their attention and passion to play the games. As a result, the game objects are distributed in entirely different locations for each player in each access to the game.

However, for fair competition in a video game, for example, each competitor must be given a randomly distributed game object with the same location and in the same order during the entire competition period. Such a scenario needs "**reproducible randomness**" [3]; namely the random results should be reproducible if given the same input.

One of the popular existing methods to generate unpredictable but reproducible randomness is Mersenne Twister 19937, or better known as MT19937 [2, 6]. MT19937 uses an integer seed value as an input to generate a sequence by a matrix linear recurrence that repeats itself only after $2^{19937} - 1$ (32-bit) integers have been produced. However, a hacker easily guests its initial conditions by exclusively searching all prime numbers within 0 to $2^{19937} - 1$ since the MT19937 seed input and generated sequence are integers and prime numbers. Its security is relatively low.

Due to their excellent chaotic properties, chaotic maps can be used to generate pseudorandom sequences to overcome the security issue with MT19937. As a classic example, the Logistic map [1, 5] was developed by Pierre-Francois Verhulst. This map is highly sensitive to its initial conditions, which are real numbers with ideally infinite precision.

This study analyzes the logistic map and compares it with the current most popular method of PRNG generation – more specifically, MT19937 – to analyze the viability of chaotic maps for PRNG in game design.

## RELATED WORKS

### MT19937

As the most popular pseudorandom number generator, the Mersenne Twister (MT) [4] was developed by Takuji Nishimura and Makoto Matsumoto in 1997. It uses an integer seed value to generate a sequence by a matrix linear recurrence that only repeats itself after $2^{19937}$ - 1 (32-bit) integers have been produced. Hence, this generator is named as 19937. The Mersenne prime numbers are the prime numbers whose value is 1 less than a power of 2. For example, a Mersenne prime 7 is $2^3 - 1$.

Using a matrix linear recurrence, it is a twisted generalized feedback shift register of rational standard form with state bit reflection and tempering. MT19937 generates numbers in the range of 0 to $2^n - 1$ for a word length of n.

### Logistic Map

A chaotic sequence has its roots in a branch of mathematics called chaos theory. Chaos theory studies the underlying patterns behind dynamical systems highly sensitive to initial conditions and were once thought to be completely random. For example, the butterfly effect is a well-known underlying principle of chaos theory – where a butterfly flapping its wings in Texas can cause a tornado in Brazil. A chaotic map can reflect the butterfly effect very well, as a slight fluctuation in its initial parameters can cause the sequence to change entirely.

As an example of a chaotic map, the Logistic map [5], discovered by Pierre-Francois Verhulst, is mathematically defined in Equation (1)

$$x_{n+1} = rx_n(1 - x_n) \tag{1}$$

where the input and output $x \in [0,\ 1]$ and parameter $r \in [0,\ 4]$ .

The Logistic Map was first used to represent the ratio between the existing population and the maximum possible population of a specific demographic. It was used to simulate reproduction and starvation within the demographic. Later, its chaotic properties were picked up by computer scientists to use for multimedia encryption, fractal generation, and pseudorandom number generator (PRNG). The Logistic map was used for PRNG because of its chaotic property and high sensitivity to initial conditions. As can be seen from the bifurcation diagram of the Logistic map in Figure 1, the Logistic map is chaotic when $r \in [3.57,\ 4]$, approximately.

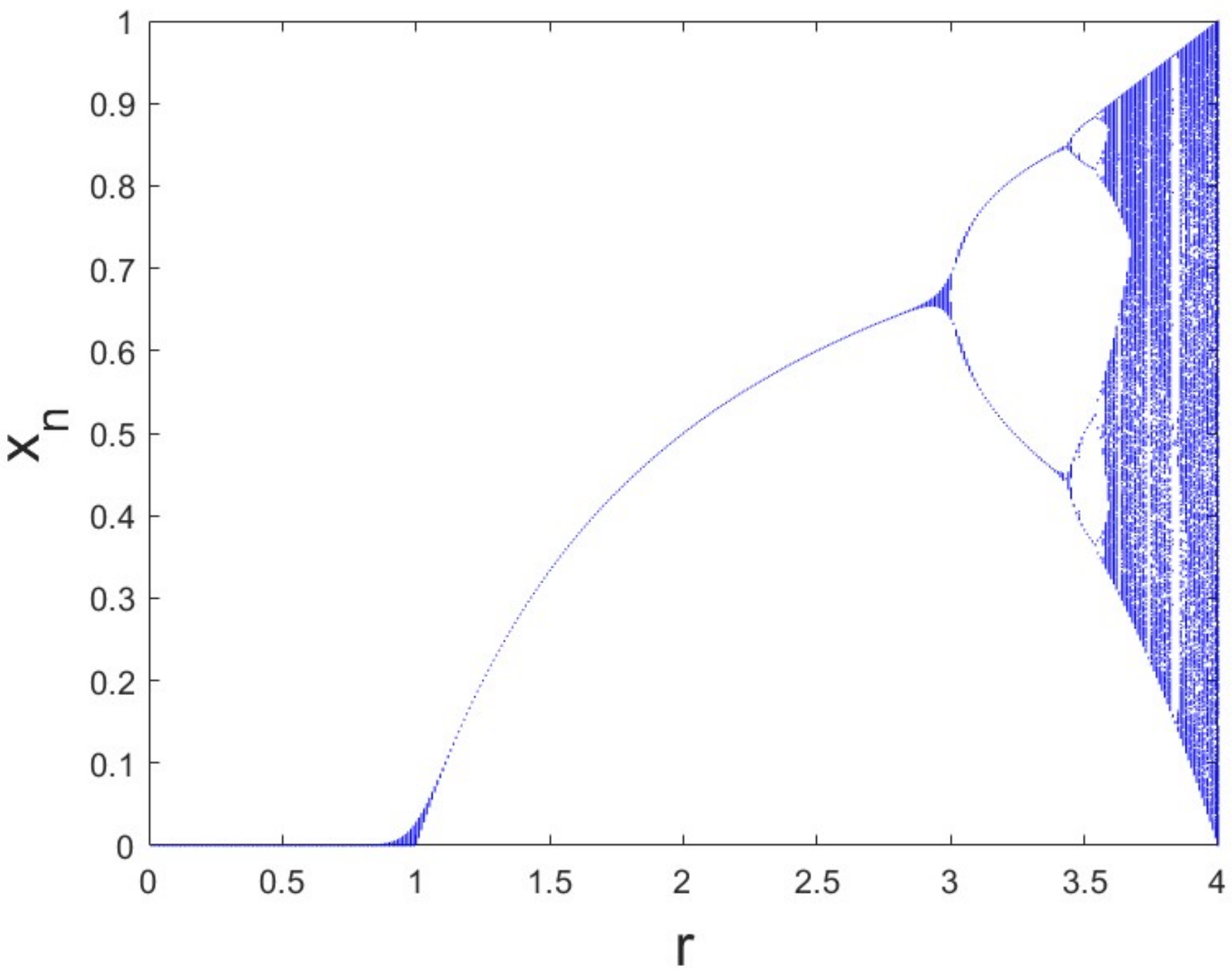

**Figure 1** Bifurcation diagram of the Logistic map. The map is more chaotic in the shaded blue areas.

**RESULTS**

**Random Sequence Generation Algorithm**

Using the Logistic map, I developed a random sequence generation algorithm, as shown in **Algorithm 1**. Using a variable array (called seed) consisting of the initial state $x_0$ and parameter $a$ as input, **Algorithm 1** can produce a random sequence with a length of $N$.

**Algorithm 1:** Random Sequence Generation Using Logistic Map

**Input:** Initial state $x_0$, parameter $r$, sequence length $N$

$x \leftarrow x_0$;

**for** $i = 1 : 30$ **do** ; `/* Ignore first 30 points */`

  $x = rx(1 - x)$

**end**

$S_1 \leftarrow x$;

**for** $i = 1 : (N - 1)$ **do** ; `/* generate random sequence */`

  $S_{i+1} = rS_i(1 - S_i)$

**end**

**Output:** random sequence $S$

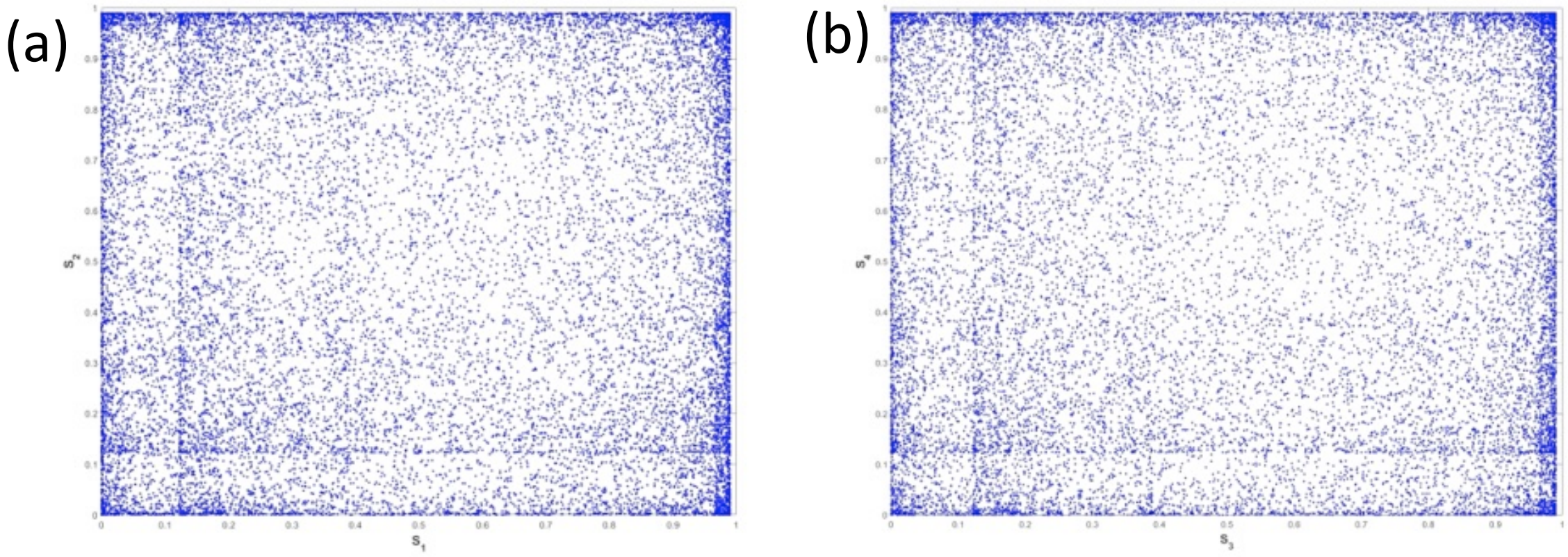

**Figure 2** Correlations of two random sequences of length 20,000 generated by the Logistic map with a tiny vibration applied to (a) initial state $x_0$ , and (b) parameter $r$. In two plots, $S_1$ and $S_3$ are generated with $r = 3.99$ and $x_0 = 0.25$; $S_2$ is generated with $r = 3.99$ and $x_0 = 0.25 + 1 \times 10^{-10}$; $S_4$ is generated with $x_0 = 0.25$ and $r = 3.99 + 1 \times 10^{-10}$.

**Sensitivity Test**

The Logistic map is very sensitive to its initial state $x_0$ and parameter $r$. Figure 2 shows the experimental results of sensitivity analysis. Algorithm 1 is used to produce random sequences when a tiny vibration ($10^{-10}$) is applied to the initial state $x_0$ (Figure 2(a)) and to the parameter $r$ (Figure 2(b)). Any tiny vibration of the initial state $x_0$ or/and parameter $r$ will lead to a completely different random sequence. For example, random sequences $S_1$ and $S_2$ are generated with the same parameter setting of $r = 3.99$ but tiny different settings of the initial state, namely $x_0 = 0.25$ for $S_1$ and $x_0 = 0.25 + 1 \times 10^{-10}$ for $S_2$. Figure 2 shows the pixel-wise correlations between sequences $S_1$ and $S_2$, and between $S_3$ and $S_4$. They are totally different sequences. These results demonstrate that the Logistic map is quite sensitive to the tiny vibration of its initial state or/and parameter.

**Comparison of Statistical Randomness**

Using MT19937 and the Logistic map as random sequence generators, each generator produces a random sequence. To test their statistical randomness of the values, I have run some statistical analysis on each generated sequence. The results are output in Table 1. All the statistical tests are run with a population size of 1000, a seed of 624 for the MT19937, an initial state $x_0 = 0.25$ and parameter value $r$ of 3.995 for the Logistic Map. As can be seen, the histograms show that the two generators can produce a random sequence whose elements have non-repeated values and are well-distributed in the entire data range of $[0,\ 1]$ as can also be seen by the slope of the LSRL of both equations in the scatter plots. However, it is noted that the standard deviation of the Logistic Map is more significant than that of MT19937, and the values of the Logistic Map are somewhat more tended toward the extremes, as shown in the scatter plot and the histogram.

**Table 1** Statistical comparison of different methods

| Statistical Tests | Logistic map | MT19937 |
|---|---|---|
| Histogram | 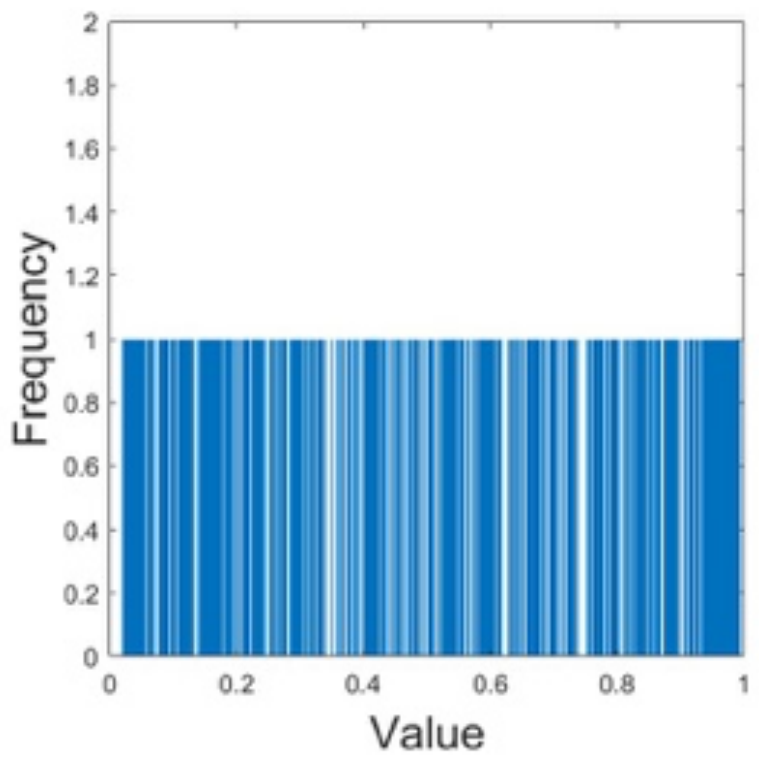 | 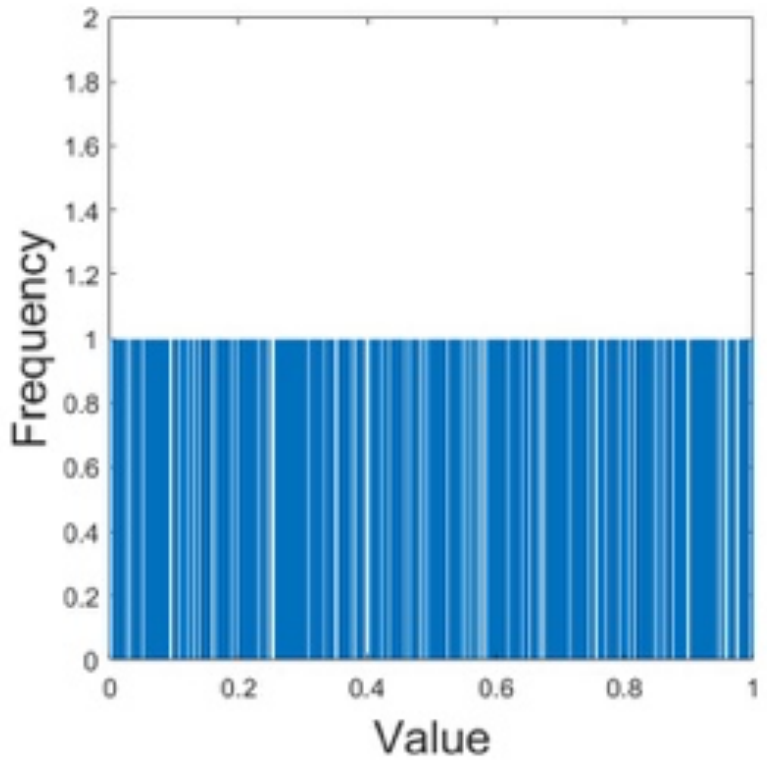 |
| Standard Deviation | 0.3364 | 0.2898 |
| Scatter Plot | 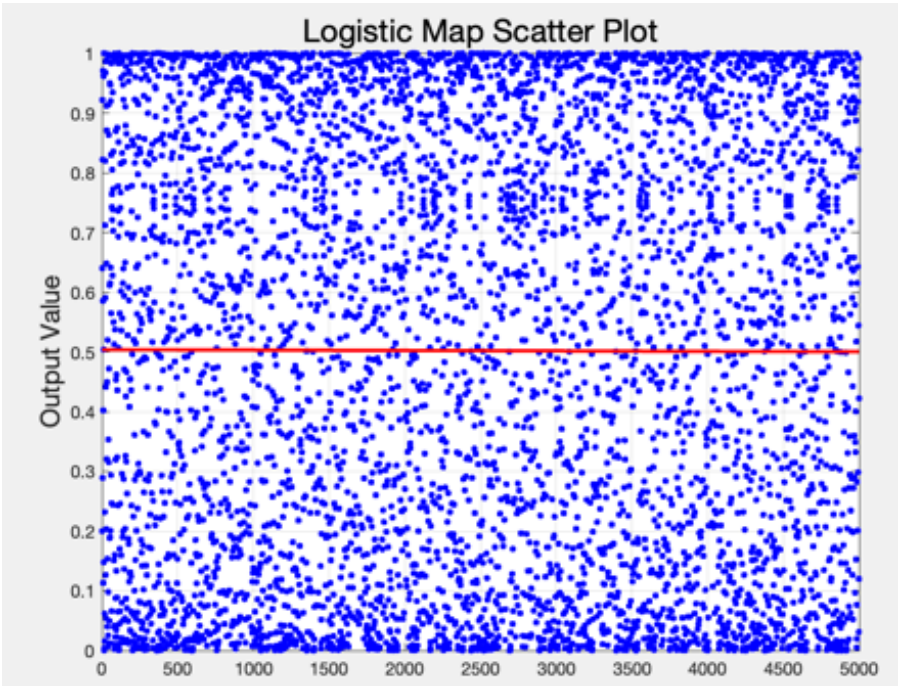 | 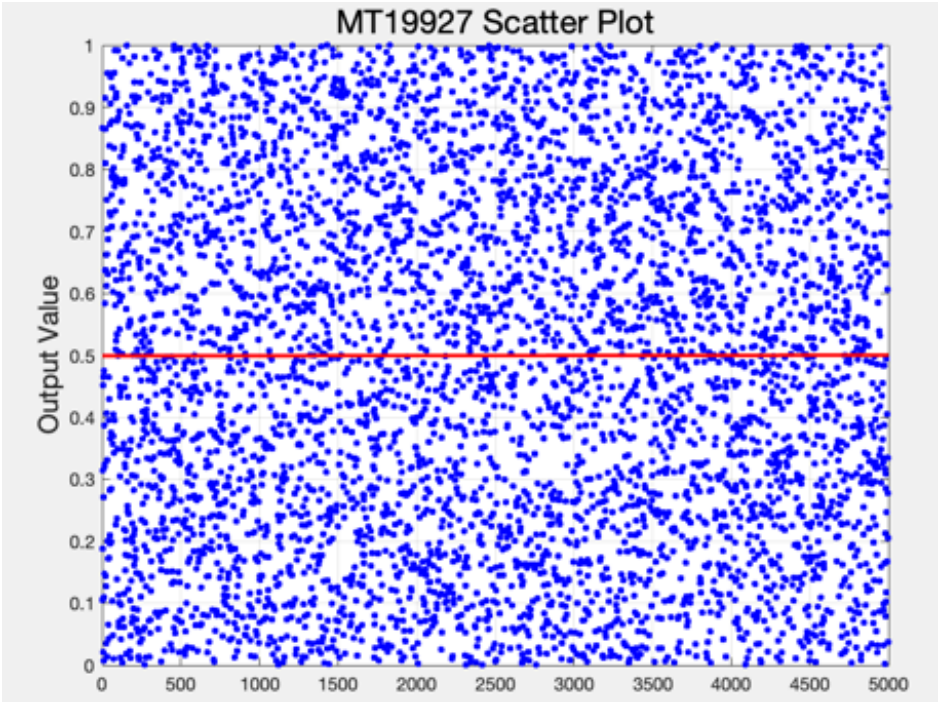 |
| LSRL Equation | 0.5038 + 0.0000x | 0.4988 + 0.0000x |

**Table 2** Comparison of the seed inputs of different methods

| Method | Seed input | Possible choices of seed input |
|---|---|---|
| MT19937 | Integer from $(0, 219937\ -\ 1)$ | $2^{19937}$ |
| Logistic map | Initial state $x_0 \in [0,\ 1]$, and parameter $r \in [3.57,\ 4]$ in the chaotic range | Theoretically Infinite |

Table 2 lists all possible choices of the seed input of these methods. MT19937 has quite limited choices of seed input because it uses an integer of 19937 bits. However, the Logistic map theoretically has infinite possible choices of seed input.

Thus, with the statistics analyzed above, the chaotic sequence is better than MT19937 for the number of different seeds available, as MT19937 only has a finite number of seeds, while the Logistic Map can have a theoretically infinite number of seeds available to use. However, the distribution is not entirely even and needs improvement.

**Game Object Random Generation Algorithm**

To demonstrate the performance of the Logistic map for random object generation in video games, I used Snake as a model platform and compared the results with MT19937. A seed variable array is first created to ensure that the Logistic map is in its chaotic range. It consists of an initial state of $x_0 \in [0,\ 1]$, and a parameter value of $r \in [3.57,\ 4]$. These two values are input into Algorithm 1 to produce a random sequence $S$ with a length of the number of pixels on the playing field (e.g. $M \times N$). A new sequence $R$ is obtained by sorting all element values of the sequence $S$

---

**Algorithm 2:** Generation of Random locations of Game Objects

---

**Input:** Initial state $x_0$, parameter $a$, screen size $M \times N$

**if** *true* **then** ; `/* for game competition */`

  $s \leftarrow x_0$

**else if** *false* **then** ; `/* when normal game play */`

  $s \leftarrow x_0 + randnoise * 10^{-12}$ ; `/* tiny noise vibration */`

**end**

$L \leftarrow M \times N$ ; `/* sequence length */`

Generate random sequence $S$ using **Algorithm 1** ($s$, $a$, and $L$) ;

$R \leftarrow sort(S)$;

**for** $i = 1 : L$ **do** ; `/* generate random locations */`

  $Z_i \leftarrow index(S == R_i)$ ; `/* random index sequence */`

  $X_i \leftarrow \mod \frac{Z_i}{M}$;

  $Y_i \leftarrow \lfloor \frac{Z_i}{M} \rfloor$;

**end**

**Output:** random locations $(X, Y)$

---

in an increasing order. Then, the new sequence $R$ is compared to the original sequence $S$, and the index of the element in the original sequence $S$ is output as another array called index sequence $Z$. Afterwards, the generated index sequence $Z$ is used to determine which location the game object is on the screen, counting from left to right, top to bottom. The $n^{th}$ term of the index sequence $Z$ is taken for the $n^{th}$ game object. For the x location of the game object, I take the modulus of the corresponding index term in $Z$ divided by the length of the playing field (e.g. $M$); and for the y location of the game object, I obtain the floor integer value of the index term in $Z$ divided by the playing field size (e.g. $M$).

**Table 3** Simulated Competition Scenario

| | **Logistic Map** | **MT19937** |
|---|---|---|
| **Game 1** | 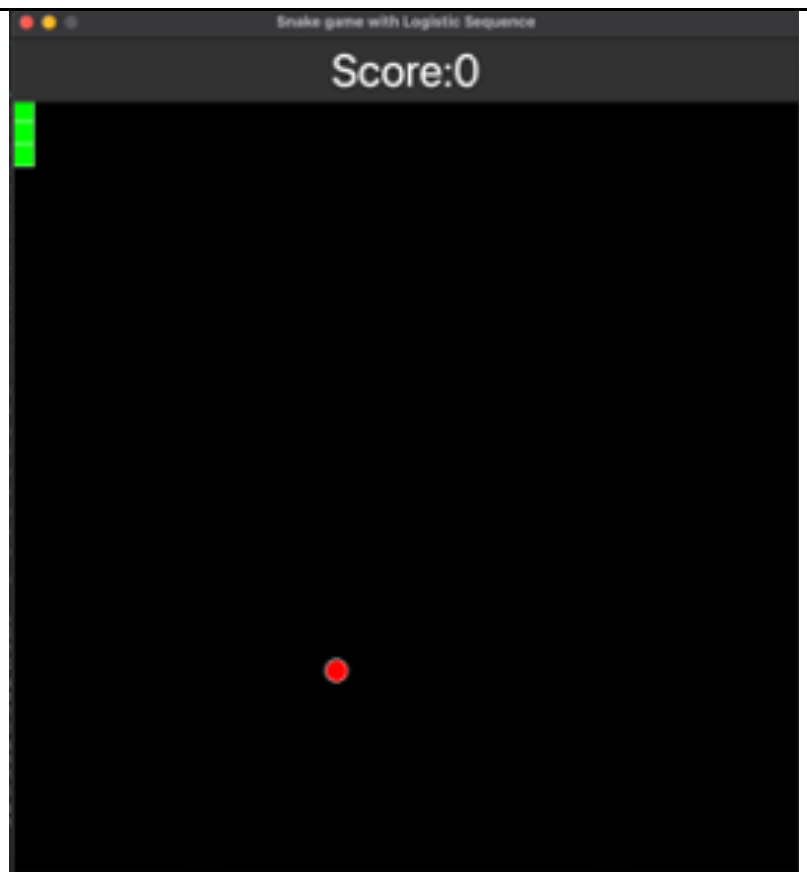  | 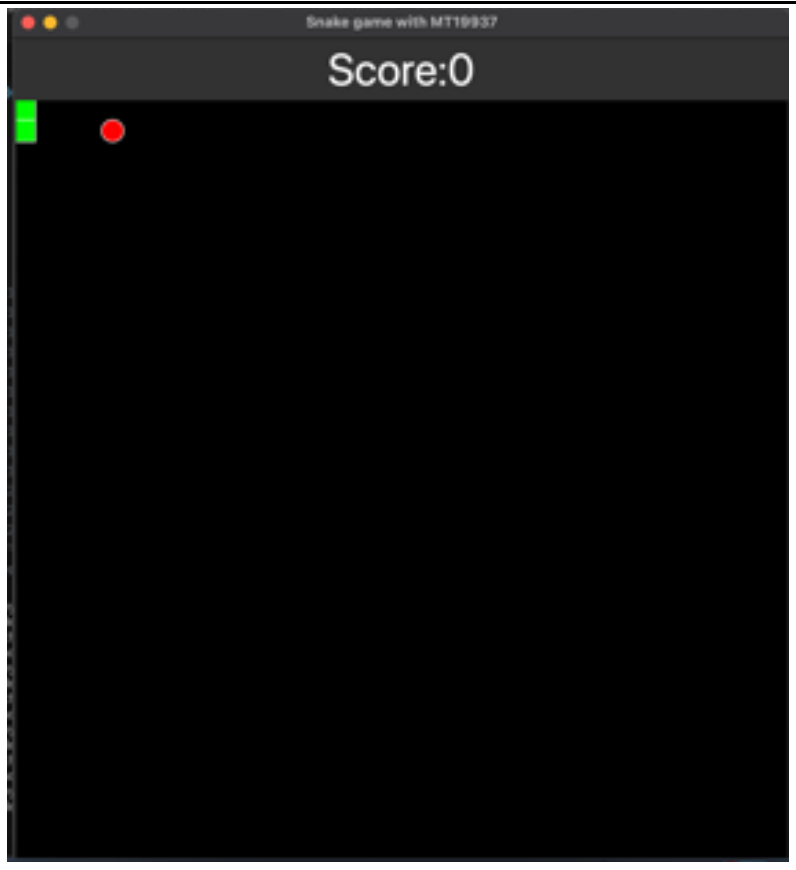  |
| **Game 2** | 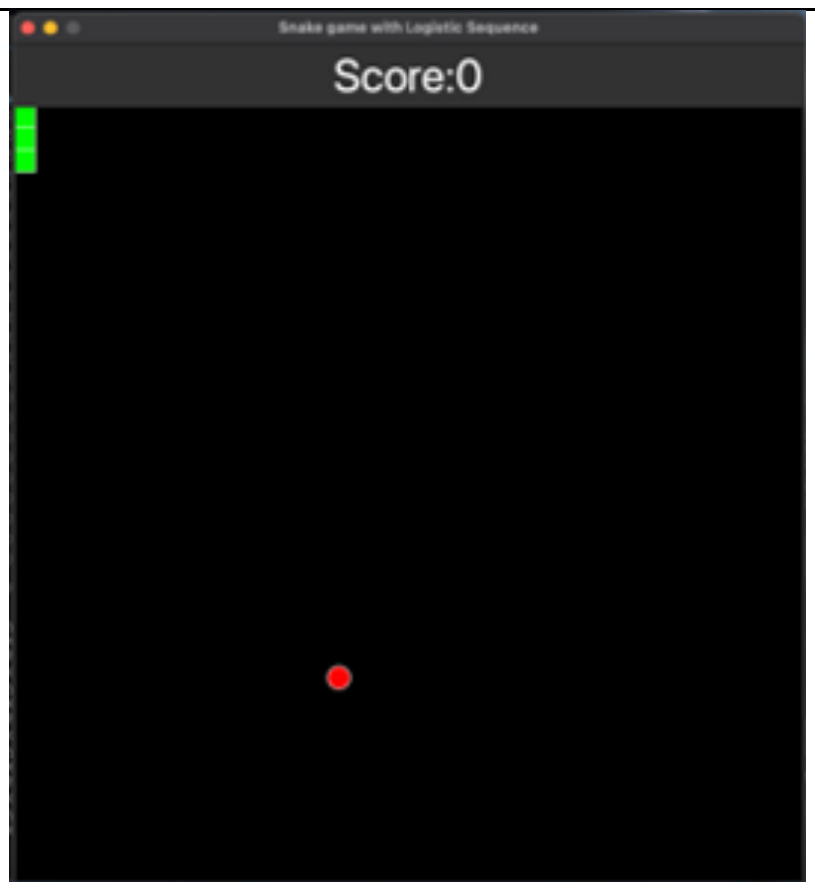  | 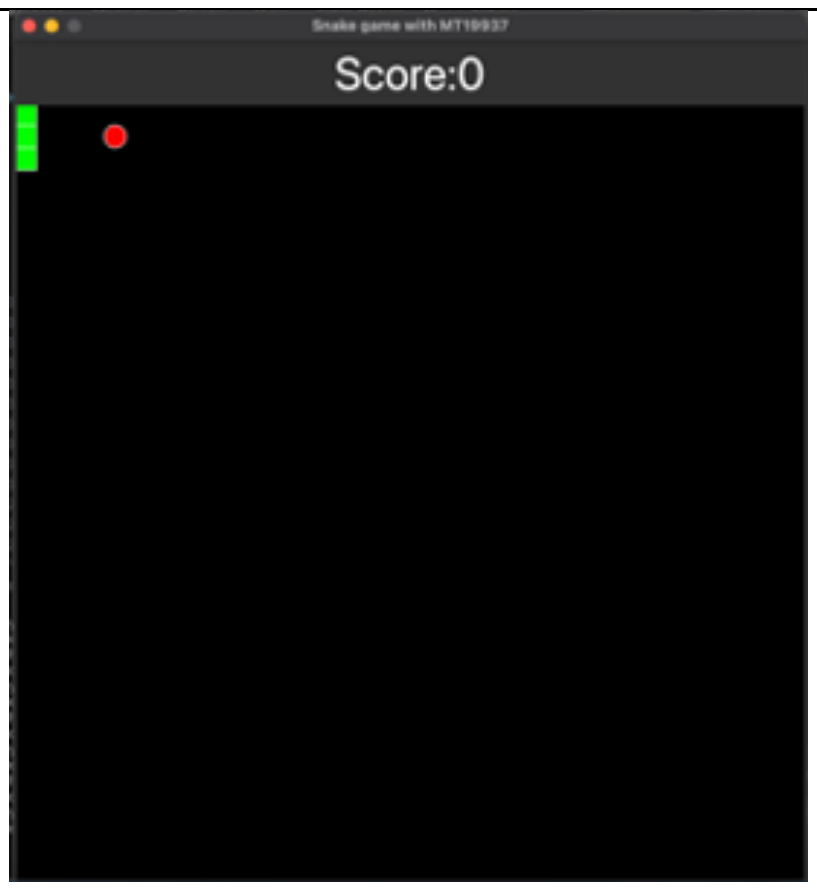  |

In such a way, the x and y coordinates for the game object are obtained. The initial term is also run through a Logistic sequence 50 times, then plugged back into the sequence due to the first few values having a strong correlation. Algorithm 2 shows the detailed procedures for generating random locations of the game objects.

**Reproducible Random Seed Generation in Snake**

By applying the two random number generation algorithms above to a video game of Snake, food generation can be randomly distributed but reproducible using the same seed. Through some testing and online evidence, the sequence is consistently sensitive to up to 12 decimal places in both the parameter and the initial value. With this number of combinations for seeds that are usable, the security of this random sequence generation should be good enough for competition purposes. The Python file sizes were the exact same for both algorithms, and the structure of code was similar. Table 3 shows the simulated competition scenarios for Logistic Map and the MT19937 generated games. The seeds can be reproducibly generated at the same location for repetitious games. In this way, the player can focus on drilling at certain locations. Furthermore, the seeds can be generated in the same sequence for different players for a fair competition.

**CONCLUSIONS**

To address limitations of existing random generation methods that have limited seed input, this study analyzed the use of the logistic map for game development. I have analyzed the properties of the Logistic map and developed a generation algorithm for pseudorandom sequences and a generation algorithm for random locations of game objects. To investigate the application of the Logistic Map and MT19937, experiments on the game of Snake demonstrated its feasibility in game development. Reproducible randomness can be realized with the proposed chaotic map algorithm. It is noted that the randomness generated with the Logistic map is more located on the edges. Further research will be required to achieve a more even generation of random numbers for chaotic maps to be viable in game development instead of using the existing methods.